\documentclass[a4paper]{article}

\usepackage{amsmath,comment,url}
\usepackage{mathtools}
\usepackage{amsthm}
\usepackage{amssymb}
\usepackage{graphicx}
\usepackage{epstopdf}

\begin{document}

\begin{center}

\section*{\LARGE \bf{Hamiltonian models for the propagation of long gravity waves, higher-order KdV-type equations and integrability  }}

\vskip1cm

{\large \bf  Rossen I. Ivanov }
\vskip1cm

\hskip-.3cm
\begin{tabular}{c}
\\
{\it School of Mathematics and Statistics, TU Dublin, City Campus, }\\ {\it Grangegorman Lower, Dublin D07 ADY7, Ireland} \\
\\
\\
\\
{\it e-mail: rossen.ivanov@tudublin.ie  }
\\
\end{tabular}
\end{center}

\vskip0.5cm

\input epsf





\begin{abstract}
A single incompressible, inviscid, irrotational fluid medium bounded above by a free surface is considered. The Hamiltonian of the system is expressed in terms of the so-called Dirichlet-Neumann operators. The equations for the surface waves are presented in Hamiltonian form. Specific scaling of the variables is selected which leads to a KdV approximation with higher order nonlinearities and dispersion (higher-order KdV-type equation, or HKdV). The HKdV is related to the known integrable PDEs with an explicit nonlinear and nonlocal transformation.  
\end{abstract}

{\bf Mathematics Subject Classification (2010):} 
76B15 (Water waves, gravity waves; dispersion and scattering, nonlinear interaction)
35Q35 (PDEs in connection with fluid mechanics), 37K10 (Completely integrable infinite-dimensional Hamiltonian and Lagrangian systems, integration methods, integrability tests, integrable hierarchies (KdV, KP, Toda, etc.))
\\
\\
{\bf Keywords:} Dirichlet-Neumann Operators, Water waves, Solitons, KdV equation, Kaup-Kuperschmidt equation, Sawada-Kotera equation, KdV hierarchy.



\section{Introduction}
In 1968 V. E. Zakharov in his work \cite{Zakharov} demonstrated that the equations for the surface waves of a deep inviscid irrotational water have a canonical Hamiltonian formulation. This result has been extended to many other situations, like long-wave models for finite depth and flat bottom \cite{CraigGroves,CraigSulem,ZhiSig}, short and intermediate wavelength water waves \cite{CCRI,I23}, internal waves between layers of different density \cite{Lan,CraigGuyenneKalisch} as well as waves with added shear with constant vorticity \cite{CompelliIvanov1,CompelliIvanov2, CI, CIM,CIP,CuIv}. We provide a detailed derivation of the higher-order KdV-type model from the Hamiltonian formulation. The Hamiltonian is expressed with the Dirichlet-Neumann Operators. These operators have known asymptotic expansions with respect to certain scale parameters which makes them convenient for the derivation of asymptotic PDE models with respect to these scale parameters. 
We establish also the relation between the obtained HKdV model and the three known integrable PDEs with nonlinear and dispersive terms of the same types.

\section{General setup of the problem and governing equations}
\begin{figure}[ht!]
\begin{center}
\fbox{\includegraphics[totalheight=0.23\textheight]{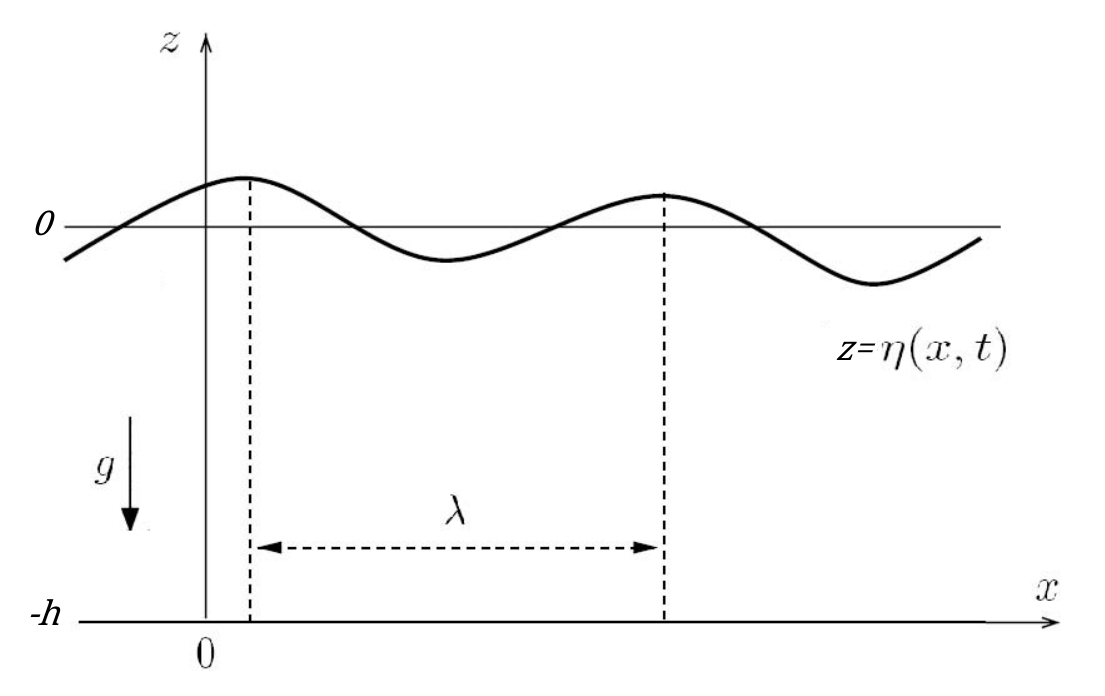}}
\caption{The system under study.}
\label{fig:thesisfigure_systemG}
\end{center}
\end{figure}

An inviscid, incompressible and irrotational fluid layer of uniform density with a free surface and flat bottom is considered, as shown in Fig. \ref{fig:thesisfigure_systemG}.

The mean surface level is located at $z=0$ (where $z$ is the vertical coordinate) and the wave elevation is given by the function $\eta(x,t).$  Therefore we have
\begin{equation}
\int_{\mathbb{R}} \eta(x,t) \, dx =0.
\end{equation}
The flat bottom is at $z=-h.$ The body of the fluid which occupies the domain $\Omega$ is defined as
\begin{equation}
\label{Omega_1_VarBott}
\Omega:=\{(x,z)\in \mathbb{R}^2: -h < z  < \eta(x,t)\}.
\end{equation}

The subscript notation $s$ will be used to refer to evaluation on the surface $z=\eta(x,t)$ and $b,$ if necessary, will refer to evaluation on the bottom $z=-h$.

Let us introduce the velocity field ${\bf{v}}=(u,w),$ where $w$ is the vertical component. The incompressibility $u_x+w_z=0$ and the irrotationality of the flow $u_z-w_x=0$ allow the introduction of a stream function $\psi(x,z,t)$ and velocity potential $\varphi(x,z,t)$ as follows:
\begin{equation}
\label{definitionofpsi_SystemG}
\begin{dcases}
    u=\psi_z=\varphi_x\\
    w=-\psi_x=\varphi_z.
\end{dcases}
\end{equation}

\noindent In addition,  $$\Delta \varphi=0, \qquad \Delta \psi =0 $$ in $\Omega.$ This leads to 
\begin{equation}\label{div}
    |\nabla \varphi|^2 = \nabla \cdot (\varphi \nabla \varphi)= \text{div} (\varphi \nabla \varphi),
\end{equation}
where $\nabla=(\partial_x, \partial_z),$ $\Delta=\nabla^2.$

The Euler equations written in terms of the velocity potential produce the Bernoulli condition on the surface
\begin{equation}
\label{Bernoulli}
(\varphi_t)_s+\frac{1}{2}|\nabla \varphi|_s^2+g \eta=0
\end{equation}
where $g$ is the acceleration due to gravity.

There is a kinematic boundary condition on the wave surface given by
\begin{equation}
\label{KBC}
w=\eta_t +u\eta_x \quad \text{or}  \quad (\varphi_z)_s=\eta_t + (\varphi_x)_s \eta_x
\end{equation}
and on the bottom
\begin{equation}
\label{KBC_Bott}
(\varphi_z)_b=0.
\end{equation}

The equations \eqref{Bernoulli} and \eqref{KBC} suggest that the dynamics on the surface is described by two variables -- the surface elevation $\eta$ and the velocity potential $\phi=(\varphi)_s.$ In fact, it turns out that these are canonical Hamiltonian variables in the Zakharov's formulation \cite{Zakharov}, which we present in the next section.  

We make the assumption that the functions $\eta(x,t)$, $\varphi(x,z,t)$ are in the Schwartz class with respect to the $x$ variable, for all possible values of the other variables\footnote{The Schwartz class function is essentially a function $ f(x)$ such that $f(x), f'(x), f''(x), ...$  all exist everywhere on $\mathbb{R}$ and go to zero as $x\to \pm \infty$ faster than any reciprocal power of $x.$ }. 
In other words we describe the propagation of solitary waves.

For rigorous mathematical results about a single layer of fluid one could refer to the monographs \cite{Constb,Lanb}. A comprehensive survey, derivation and analysis of the nonlinear water wave models is presented in \cite{Johnson}. 

\section{Hamiltonian formulation}
The Hamiltonian of the system  \eqref{Bernoulli} -- \eqref{KBC_Bott} will be represented as the total energy of the fluid with density $\rho,$
\begin{equation}
H=\frac{1}{2}\rho \int\int_{\Omega} (u^2+w^2)dz dx+\rho g \int \int _{\Omega} z dz dx.
\end{equation}
It can be written in terms of the variables $(\eta(x,t),\varphi(x,z,t)),$ using \eqref{div}, as
\begin{equation} \label{8}
H[\eta,\varphi]= \frac{1}{2}\rho \int_{\mathbb{R}}  \int_{-h}^{\eta} \text{div} (\varphi \nabla \varphi) dz dx
+\rho g \int_{\mathbb{R}}  \int_{-h}^{\eta} z dz dx.
\end{equation}

We introduce the variable $\xi,$ which is defined to be proportional to the potential evaluated on the surface
\begin{equation}
\xi(x,t):=\rho \varphi(x,\eta(x,t),t)\equiv \rho \phi(x,t),
\end{equation}

\noindent and the Dirichlet-Neumann operator $G(\eta)$ defined by
\begin{equation}
 G(\eta)\phi =-\eta_x (\varphi_x)_s+(\varphi_z)_s =(-\eta_x, 1) \cdot (\nabla \varphi)_s=\sqrt{1+\eta_x^2}\,\,  {\bf n}_s \cdot (\nabla \varphi)_s
\end{equation}
where  ${\bf n}_s = (-\eta_x, 1)/\sqrt{1+\eta_x^2}$ is the outward-pointing unit normal vector (with respect to $\Omega$) to the wave surface.

Applying Green's Theorem (Divergence Theorem) to \eqref{8}, the Hamiltonian can be written as
\begin{equation}
\label{mainHamiltonian}
H[\eta, \xi]=\frac{1}{2\rho}\int_{\mathbb{R}} \xi G(\eta) \xi dx
+\frac{1}{2}\rho g\int_{\mathbb{R}} (\eta^2-h^2) dx .
\end{equation}
On the bottom the outward-pointing unit normal vector is
\noindent ${\bf n}_b=(0, -1)$ and   $${\bf n}_b \cdot (\nabla \varphi)_b =(0, -1) \cdot \left(( \varphi_x)_b, (\varphi_z)_b\right)=(0- (\varphi_z)_b)=0  $$ thus $  {\bf n}_b \cdot (\nabla \varphi)_b =0$ and no bottom-related terms are present in \eqref{mainHamiltonian}. Noting that the term $\int_{\mathbb{R}} h^2(x) dx$ is a constant and will not contribute to $\delta H$, we renormalize the Hamiltonian to
\begin{equation}
\label{mainHam}
H[\eta, \xi]=\frac{1}{2\rho}\int_{\mathbb{R}} \xi G(\eta) \xi dx
+\frac{1}{2}\rho g\int_{\mathbb{R}} \eta^2 dx.
\end{equation}
The variation of the Hamiltonian can be evaluated as follows. An application of Green's Theorem transforms the following expression to one which involves contributions from the surface and the bottom alone:

\begin{align}
\delta\bigg[\frac{\rho }{2} \int_{\mathbb{R}}  \int_{-h}^{\eta}  (\nabla \varphi)\cdot  (\nabla \varphi) & \, dz dx\bigg] \nonumber \\
&=\rho \int_{\mathbb{R}}  \int_{-h}^{\eta}  (\nabla \varphi)\cdot  (\nabla \delta \varphi) dz dx  +\frac{1}{2}\rho\int_{\mathbb{R}} (|  \nabla \varphi  |^2)_s \, \delta \eta \,dx \notag\\
&= \rho \int_{\mathbb{R}}  \int_{-h}^{\eta} \text{div} [(\nabla \varphi) \delta \varphi] dz dx  +\frac{1}{2}\rho\int_{\mathbb{R}} (|  \nabla \varphi  |^2)_s \, \delta \eta \,dx \notag\\
&=\rho\int_{\mathbb{R}} \big((\varphi_z)_s-(\varphi_x)_s \eta_x\big)(\delta\varphi)_s dx\notag\\
&\quad-\rho\int_{\mathbb{R}} (\varphi_z)_b(\delta\varphi)_b dx
+\frac{1}{2}\rho\int_{\mathbb{R}}(|  \nabla \varphi  |^2)_s  \, \delta \eta \,dx.
\end{align}
Due to \eqref{KBC_Bott}, the contribution from the term evaluated on the bottom vanishes, thus
\begin{align}
\delta\bigg[\frac{\rho }{2} \int_{\mathbb{R}}  \int_{-h}^{\eta} |  \nabla \varphi  |^2 dz dx\bigg]&=\rho\int_{\mathbb{R}} \big((\varphi_z)_s-(\varphi_x)_s \eta_x\big)(\delta\varphi)_s dx\notag\\
& +\frac{1}{2}\rho\int_{\mathbb{R}} (|  \nabla \varphi  |^2)_s \, \delta \eta \,dx.
\end{align}
\noindent Clearly,
\begin{align}
\delta\bigg[\rho g\int_{\mathbb{R}} \eta^2 dx\bigg]&=2\rho g\int_{\mathbb{R}} \eta\delta\eta dx.
\end{align}
\noindent Noting that the variation of the potential on the wave surface is given as
\begin{equation}
(\delta\varphi)_s=\delta\phi-(\varphi_z)_s\delta\eta,
\end{equation}
where
\begin{equation}
\phi(x,t) :=\varphi(x,\eta(x,t),t),
\end{equation}
we write
\begin{multline}
\delta H=\rho\int_{\mathbb{R}} \big((\varphi_z)_s-(\varphi_x)_s \eta_x\big)\big(\delta\phi-(\varphi_z)_s\delta\eta\big) dx
+\frac{1}{2}\rho\int_{\mathbb{R}}  |\nabla \varphi|_s^2\, \delta \eta \,dx
+\rho g\int_{\mathbb{R}} \eta\delta\eta dx. \nonumber
\end{multline}

Evaluating $\delta H/ \delta\xi$ we remember that $\rho\delta\phi=\delta\xi$ and therefore

\begin{equation}
\frac{\delta H}{\delta \xi}=(\varphi_z)_s-(\varphi_x)_s \eta_x=\eta_t
\end{equation}
\noindent due to (\ref{KBC}).  Next we compute
\begin{equation}
\frac{\delta H}{\delta \eta}= - \rho \big((\varphi_z)_s-(\varphi_x)_s \eta_x\big)(\varphi_z)_s    +\frac{1}{2} \rho |\nabla \varphi|_s^2  +\rho g \eta.
\end{equation}

\noindent Noting that, using the kinematic boundary condition (\ref{KBC}),
\begin{align}
 -  \big((\varphi_z)_s-(\varphi_x)_s \eta_x\big)(\varphi_z)_s
 & = - \eta_t  (\varphi_z)_s
\end{align}
we write
\begin{equation}
\frac{\delta H}{\delta \eta}= - \rho \eta_t(\varphi_z)_s    +\frac{1}{2}\rho  |\nabla \varphi|_s^2  +\rho g \eta.
\end{equation}
Recall that $$ \xi_t = \rho((\varphi_t)_s   +(\varphi_z)_s \eta_t), $$
and so
\begin{align}
\frac{\delta H}{\delta \eta}&= - \xi_t + \rho  \left( (\varphi_t)_s   +\frac{1}{2}  |\nabla \varphi|_s^2+ g \eta \right)= - \xi_t
\end{align}
by the virtue of the Bernoulli equation (\ref{Bernoulli}). Thus we have canonical equations of motion
\begin{equation}
\label{EOM}
    \xi_t= -\frac{\delta H}{\delta \eta}, \qquad
    \eta_t= \frac{\delta H}{\delta \xi}
\end{equation}
\noindent where the Hamiltonian is given by \eqref{mainHam}. Introducing the variable $\mathfrak{u}=\xi_x,$ which is proportional to the horizontal velocity along the free surface, by changing the variable, we can represent \eqref{EOM} in the form
\begin{equation}
\label{EOM1}
    \mathfrak{u}_t= -\left(\frac{\delta H}{\delta \eta}\right)_x, \qquad
    \eta_t= - \left(\frac{\delta H}{\delta \mathfrak{u}}\right)_x,
\end{equation}
which can also be expressed in the matrix form
\begin{equation}\label{op_E}
\begin{pmatrix} \mathfrak{u} \\ \eta \end{pmatrix}  _t  =  -  \begin{pmatrix} 0  & 1\\
1  & 0 \end{pmatrix} \begin{pmatrix} \frac{\delta {H}}{\delta \mathfrak{u}} \\ \frac{\delta {H}}{\delta \eta}\end{pmatrix}  _x.
\end{equation}

The Hamiltonian can be expressed through the canonical variables $ \mathfrak{u}, \eta$ by using the properties of the Dirichlet-Neumann operator, which are introduced in the next section.  Thus we have a formulation of the problem involving the surface variables alone.

\section{The Dirichlet-Neumann operator}
We begin this section with some basic properties of the Dirichlet-Neumann operator. The details can be found in \cite{CraigGroves,CraigSulem,IKI23}. The operator can be expanded as
\begin{equation}
G(\eta)=\sum_{j=0}^{\infty} G^{(j)}(\eta)
\end{equation}
where $ G^{(j)}(\eta)\sim (\eta/h)^j.$  The surface waves are assumed to be small, relative to the fluid depth, that is, $\varepsilon=|\eta_{\mathrm{max}}|/h\ll 1$ is a small parameter, and hence one can expand with respect to $|\eta/h |\ll 1 $ as follows: \begin{align}
    G^{(0)}=&D\tanh(hD) , \\
    G^{(1)}=&D\eta D -G^{(0)}\eta G^{(0)}, \\
    G^{(2)}=&-\frac{1}{2 }(D^2\eta^2 G^{(0)} +   G^{(0)} \eta^2 D^2- 2G^{(0)}\eta G^{(0)}\eta G^{(0)}   ),     \ldots .
\end{align}

The operator $D=-i\partial/\partial x$ has the eigenvalue $k=2\pi/\lambda$ for any given wavelength $\lambda$, when acting on monochromatic plane wave solutions proportional to $e^{ik(x-c(k)t)}.$ In the long-wave regime the parameter $\delta=h/\lambda \ll 1$ is assumed to be small and since $hD$ has an eigenvalue $$hk=\frac{2\pi h}{\lambda} \ll 1 $$ thus $h k$ is small as well, and one can formally expand in powers of $h D$ (which are of order $\delta$). As a matter of fact, the equations for a single layer of fluid could be written in terms of non-dimensional variables, see for example \cite{CJ08}. Then the quantities $h,g$ and $c$ are simply equal to one. In our considerations however we keep track of these quantities explicitly and keep in mind that all they are of order one.
The magnitude of the terms is therefore labeled explicitly by the scale parameters $\varepsilon$ and $\delta$. In the long-wave and small-amplitude regime, $hD \sim \delta \ll 1,$ (that is, $h D$ is of order $\delta$). 
\noindent Using the expansion 
\begin{align}
\tanh(hD)&=hD-\frac{1}{3}h^3D^3+\frac{2}{15}h^5 D^5+\mathcal{O}((hD)^7)\notag\\
\end{align}
and introducing explicitly the scale parameters, we obtain
\begin{equation}
\begin{split}
G(\eta)&= \delta^2 D( h + \varepsilon \eta )D -\delta^4 D^2\left[\frac{1}{3} h ^3+\varepsilon h^2 \eta \right] D^2  +  \delta^6 \frac{2}{15}h^5 D^6 \\
& \phantom{*****************************}+\mathcal{O}(\delta^8, 
\varepsilon \delta^6, \varepsilon^2 \delta^4) \label{DN_1}
\end{split}
\end{equation}

In what follows we continue by considering the so-called Boussinesq-type approximation. In essence, this approximation further assumes $\delta^2 \sim \varepsilon,$  $\xi \sim \delta,$  where the symbol $\sim$ means that the quantities are of the same order. The Boussinesq-type equations describe waves traveling simultaneously in opposing directions. 

In the leading order of the scale parameters (that is, keeping only the lowest order $\delta^2$ in \eqref{DN_1}), the operator \eqref{DN_1} is $G^{(00)}=hD^2$ and the Hamiltonian \eqref{mainHam} is therefore
\begin{equation}
\label{Ham0}
H^{(2)}[ \mathfrak{u}, \eta]=\frac{1}{2}\int_{\mathbb{R}}    \frac{h}{\rho} \mathfrak{u}^2 \, dx
+\frac{1}{2}\int_{\mathbb{R}} \rho g \eta^2 \, dx= \frac{1}{2}\int_{\mathbb{R}} Q^T {\bf A} Q \,dx.
\end{equation}
It can be represented as a quadratic form for $Q:=(\mathfrak{u},\eta)^T$ with a matrix
\begin{equation}
\label{A}
{\bf A}:= \begin{pmatrix} \frac{h}{\rho}  & 0\\
0  & \rho g \end{pmatrix}.
\end{equation}
The vector $Q$ is $2$-dimensional,
$$Q:=(  \mathfrak{u},\eta, )^T\equiv (Q_1,Q_2)^T$$ and the equations \eqref{op_E} under these assumptions are
\begin{equation}\label{Qeq0}
Q_t=-{\bf J} {\bf A} Q_x ,  \quad {\bf J}:= \begin{pmatrix} 0 & 1\\
1  & 0 \end{pmatrix}.
\end{equation}
The diagonalization of the matrix ${\bf J}{\bf A}$ is given by ${\bf J}{\bf A}=\mathbf{ VCV}^{-1}$ for 
\begin{equation} \label{c1c2}
 \mathbf{C}=\text{diag}(c_1,c_2)=\text{diag}(\sqrt{gh}, -\sqrt{gh}),   
\end{equation}
where $c_1$ and the $c_2$ can be regarded as the speeds of the right- and left-moving waves, as we will now see, and
\begin{equation}\label{Qeq01}
\mathbf{V}:= \begin{pmatrix} \rho \sqrt{\frac{g}{h}} &-\rho \sqrt{\frac{g}{h}} \\
1  & 1 \end{pmatrix}.
\end{equation}
We introduce a new variable $Z=(Z_1, Z_2)^T,$ such that $Q=\mathbf{V}Z.$ Then the equations \eqref{Qeq0} become
\begin{equation}
\label{Z0eq}
Z_t+ C Z_x=0, \, \, \text{or} \, \, (Z_i)_t+c_i (Z_i)_x=0.
\end{equation}

Thus the $Z_i=Z_i(x-c_i t)$ in this approximation are functions of the corresponding characteristic variables. These functions are localised disturbances (waves) propagating with speeds $c_i.$  We refer to the $Z_i$ as ''propagation modes''. Given the fact that all propagation speeds are different, the disturbances, (or propagation modes) $Z_i$ move with different, opposite speeds. It is reasonable therefore to make the assumption that their interaction is negligible after a certain period in time.  This means, that in the higher order approximations of the Hamiltonian, we neglect any products $Z_i Z_j$ when $i\ne j.$ The relationship between the physical variables and the propagation modes $Q=\mathbf{V}Z,$ where $\mathbf{V}$ is given in \eqref{Qeq01}, can be written explicitly in the form
\begin{equation}
\label{Q2Za}
\begin{split}
Q_1&=\mathfrak{u}=V_{11}Z_1+V_{12}Z_2=\rho \sqrt{\frac{g}{h}}(Z_1-Z_2),  \\
Q_2&= \eta=Z_1+Z_2.
\end{split}
\end{equation}
As a ''reference '' variable we take $\eta=Z_i,$  this is the elevation of the wave propagating with wave speed $c_i,$ where $i=1$ or $i=2.$ From now on we do not write explicitly the index $i,$ that is, we consider the propagation of only one of the two modes, $\eta=Z,$ moving with speed $c.$ In other words the other propagation mode is considered being identically zero. This is possible, since the interaction between both modes is neglected and modes propagate separately - in this case in opposite directions. Then eq. \eqref{Q2Za} becomes simply
\begin{equation}
\label{Q2Za1}
\begin{split}
Q_1&=\mathfrak{u}=V_{1}Z,   \\
Q_2&= \eta=Z.
\end{split}
\end{equation} 
In order to take into account nonlinear terms, we expand the Hamiltonian (\ref{mainHam}) in the scale parameter $\varepsilon,$ taking into account the assumptions of the Boussinesq approximation. Using the expansion for the corresponding Dirichlet-Neumann operator \eqref{DN_1}, keeping only terms of order $\varepsilon ^4$ we obtain:
\begin{equation}
H[Q]=\varepsilon^2  H^{(2)} +\varepsilon^3 H^{(3)}[Q]  +\varepsilon^4 H^{(4)}[Q]+\mathcal{O}(\varepsilon^5),
\end{equation} where
\begin{align}
 H^{(3)}[Q]=&-\frac{1}{2}\int_{\mathbb{R}}\frac{h^3}{3\rho} \mathfrak{u}_x^2 \, dx  +\frac{1}{2}\int_{\mathbb{R}}\frac{1}{\rho}\eta \mathfrak{u}^2 \, dx, \\
 H^{(4)}[Q]=&\frac{1}{2}\int_{\mathbb{R}}\frac{2h^5}{15\rho}
 \mathfrak{u}_{xx}^2 \, dx -
 \frac{1}{2}\int_{\mathbb{R}}\frac{h^2}{\rho} \eta
 \mathfrak{u}_{x}^2 \, dx .
 \end{align}
Taking into account the fact that the variables $\eta, \mathfrak{u}$ are both of order $\varepsilon,$  the equations of motion \eqref{op_E} are
\begin{equation}
\label{EOM2Za}
\begin{split}
\mathfrak{u}_t&=-\rho g \eta_x-\frac{\varepsilon}{\rho}  \mathfrak{u} \mathfrak{u}_x  + \varepsilon^2 \frac{h^2}{\rho}  \mathfrak{u}_x\mathfrak{u}_{xx}     , \\
\eta_t&= -\frac{h}{\rho} \mathfrak{u}_x-\frac{\varepsilon h^3}{3\rho} \mathfrak{u}_{xxx}-\frac{\varepsilon}{\rho}(\eta \mathfrak{u})_x
-\varepsilon^2 \frac{2h^5}{15 \rho} u_{xxxxx}-\varepsilon^2 \frac{h^2}{\rho} (\eta \mathfrak{u}_x )_{xx}.
\end{split}
\end{equation}
Now, our aim is to describe the time evolution of $\eta=Z$ with a single equation. To this end we wish to extend the linear relation $\mathfrak{u}=V_{1}Z  $ in \eqref{Q2Za1} to a more complex one, which is suggested by the form of the nonlinearities in $H^{(3)}, $ $H^{(4)} $ and the equations,
\begin{equation}
\label{Q2ZaNL}
\begin{split}
Q_1=\mathfrak{u}=&V_{1}Z + \varepsilon (\alpha Z^2 +  \beta  Z_{xx}) +\varepsilon^2( \gamma Z^3 + \mu Z Z_{xx} + \nu Z_x^2 + \theta Z_{xxxx})      , \\
Q_2= \eta =&Z,
\end{split}
\end{equation} where $\alpha,\beta, \gamma, \mu,\nu, \theta$ are yet unknown constant coefficients.  This relation is in fact an algebraic-differential constraint between the two Hamiltonian variables
$\mathfrak{u}$ and $\eta$, which effectively reduces twice the phase space of the Hamiltonian system.   

The time derivative of $\mathfrak{u}$ according to \eqref{Q2ZaNL} is $$\mathfrak{u}_t=\big(V_{1} + \varepsilon ( 2 \alpha Z +  \beta \partial_x^2) + \varepsilon^2(3\gamma Z^2+ \mu Z_{xx}+\mu Z \partial_x ^2 + 2\nu Z_x \partial_x+\theta \partial_x^4  )\big)  Z_{t}=:\tilde{V} Z_{t},$$
where $$\tilde{V}:=V_{1} + \varepsilon ( 2 \alpha Z +  \beta \partial_x^2) + \varepsilon^2(3\gamma Z^2+ \mu Z_{xx}+\mu Z \partial_x ^2 + 2\nu Z_x \partial_x+\theta \partial_x^4  )$$ is a self-adjoined differential operator.

Inserting \eqref{Q2ZaNL} in \eqref{EOM2Za} leads to a system which involves only the $Z$ variable and the unknown constants $\alpha, \beta, ...$ of the form
\begin{equation}
\label{EOM2Zb}
\begin{split}
\mathfrak{u}_t&\equiv \tilde{V} Z_{t}= f_1[Z], \\
\eta_t&\equiv Z_{t}= f_2[Z],
\end{split}
\end{equation} where
\begin{equation}
\label{f12Z}
\begin{split}
f_1[Z]=&-\rho g Z_{x}-\frac{\varepsilon}{\rho}V_{1}^2 Z Z_{x}\\
&-\varepsilon^2 \frac{3 \alpha V_1}{\rho} Z^2 Z_x + \varepsilon^2 
\frac{h^2 V_1^2 - \beta V_1}{\rho}Z_x Z_{xx}-\varepsilon^2 \frac{\beta V_1}{\rho} Z Z_{xxx} + \mathcal{O}(\varepsilon^3), \\
f_2[Z]=&  -\frac{h}{\rho} V_{1}Z_{x}-\varepsilon \frac{2}{\rho}\left(h \alpha +V_{1} \right) Z Z_{x} -\frac{\varepsilon h}{\rho} \left( \beta+ \frac{h^2}{3} V_{1} \right) Z_{xxx} \\
&-\varepsilon^2\left( \frac{\theta h}{\rho} + \frac{h^3\beta}{3 \rho} + \frac{2h^5 V_1}{15 \rho}\right) Z_{5x} - \varepsilon^2 \frac{3(h\gamma+\alpha)}{\rho} Z^2 Z_x \\
& - \varepsilon^2 \frac{h \mu + 2h \nu +2\alpha h^3 +\beta+3h^2 V_1}{\rho}Z_x Z_{xx}\\
&-\varepsilon^2 \left(\frac{h\mu + \beta+ h^2 V_1}{\rho}+ \frac{2\alpha h^3}{3 \rho}  \right)Z Z_{xxx}+ \mathcal{O}(\varepsilon^3).
\end{split}
\end{equation}

The equations \eqref{EOM2Zb} are compatible iff $f_1[Z]\equiv \tilde{V} f_2[Z].$ This leads to a lengthy expression for $\tilde{V} f_2[Z]$, which can be truncated up to the terms of order $\varepsilon^2.$ The comparison with $f_1[Z]$ in \eqref{f12Z} gives rise to equations, generated by matching the coefficients of the like terms. This enables the determination of the unknown constants as follows:
 \begin{align}\label{aibi}
 Z Z_x \, \text{term} \, &\rightarrow \, \alpha=-\frac{1}{4h}V_{1}, \\
 Z_{xxx } \, \text{term} \, &\rightarrow \,  \beta =-\frac{h^2}{6}V_{1}, \\
  Z^2 Z_x \, \text{term} \, &\rightarrow \, \gamma=-\frac{1}{8h^2} V_{1}, \\
   Z_{xxxxx} \, \text{term} \, &\rightarrow \, \theta=-\frac{h^4}{40}V_{1}, \\
    Z Z_{xxx} \, \text{term} \, &\rightarrow \, \mu=-\frac{h}{4}V_{1}, \\
     Z_x Z_{xx} \, \text{term} \, &\rightarrow \, \nu=-\frac{9h}{16}V_{1}. 
     \label{nu}
\end{align}

From \eqref{EOM2Zb}, the equation describing the evolution of the propagating mode $Z$ takes the form $Z_t-f_2[Z]=0.$ From \eqref{f12Z} using \eqref{aibi} -- \eqref{nu}, this can be expressed as
\begin{align}   
Z_{t}+&\frac{h V_{1}}{\rho} Z_{x}+\varepsilon \frac{h^3 V_{1}}{6\rho}Z_{xxx}+\varepsilon \frac{3V_{1}}{2\rho} Z Z_{x}+\varepsilon^2 \frac{19}{360} \frac{h^5 V_1}{\rho} Z_{5x} \nonumber\\
&-\varepsilon^2 \frac{3}{8} \frac{V_1}{h\rho} Z^2 Z_x +
\varepsilon^2 \frac{h^2 V_1}{\rho}\left( \frac{23}{24} Z_x Z_{xx}+\frac{5}{12} Z Z_{xxx}\right)=0. 
 \label{eqZi}
\end{align}
Taking into account the relations $c=\frac{hV_1}{\rho}=\pm \sqrt{gh}$ given by \eqref{c1c2}, \eqref{Qeq01} and  $\eta=Z,$ we have
\begin{align}
    \eta_t +&c\eta_x +\varepsilon \frac{c h^2 }{6}\eta_{xxx}+\varepsilon \frac{3c}{2h} \eta \eta_x \nonumber\\
    &+ \varepsilon^2 \frac{19 ch^4}{360}\eta_{5x} -\varepsilon^2 \frac{3c}{8h^2}\eta^2 \eta_x +\varepsilon^2 ch
    \left( \frac{23}{24} \eta_x \eta_{xx}+\frac{5}{12} \eta \eta_{xxx}\right)=0. 
    \label{KdV_c}
\end{align}
This is a higher order KdV-type equation (HKdV). The expression for the other variable $\mathfrak{u}$ up to $\mathcal{O}(\varepsilon^2)$ is then 
\begin{equation}
\label{eq-u-Zi}
\mathfrak{u}=V_{1}\left( Z-\frac{\varepsilon}{4h} Z^2 -\frac{\varepsilon h^2}{6} Z_{xx} - \frac{\varepsilon^2}{8h^2} Z^3-
 \frac{\varepsilon^2 h}{4}Z Z_{xx} -\varepsilon^2 \frac{9h}{16}Z_x^2 -\varepsilon^2 \frac{h^4}{40}Z_{4x} \right) ,
\end{equation} equivalently, using as before $c=\frac{hV_1}{\rho}$ and  $\eta=Z,$ we have the expression 
\begin{align}
    \label{eq-u-eta}
\mathfrak{u}&=\frac{\rho c}{h} \left(\eta-\frac{\varepsilon}{4h} \eta^2 -\frac{\varepsilon h^2}{6} \eta_{xx}
- \frac{\varepsilon^2}{8h^2} \eta^3-
 \frac{\varepsilon^2 h}{4}\eta \eta_{xx} -\varepsilon^2 \frac{9h}{16}\eta _x^2 -\varepsilon^2 \frac{h^4}{40} \eta _{4x} 
\right), 
\end{align} where the wavespeed has two possible values $c=\pm \sqrt{gh},$ due to \eqref{c1c2}.

A HKdV equation for $\mathfrak{u}$ can also be derived, but its coefficients will be different.
This can be achieved for example with a similar procedure, where the reference variable is taken to be
$\mathfrak{u} \equiv V_1 Z$ and $\eta$ is expressed in terms of $Z$ by a relation of the type
$ \eta= Z+\varepsilon (\alpha' Z^2 +  \beta'  Z_{xx}) +\varepsilon^2( \gamma' Z^3 + \mu' Z Z_{xx} + \nu' Z_x^2 + \theta' Z_{xxxx}).  $ 

The equation \eqref{KdV_c} appears in a number of previous studies involving models beyond the KdV approximation, see for example \cite{DGH,Adkdv5,Mar,ZhiSig}. This equation in general is not integrable, its relation to integrable equations with the same type of nonlinear and dispersive terms will be established in the next section.

\section{ Near-identity transformation and relation to integrable equations}

In this section we establish a relation between two HKdV equations of the type \eqref{KdV_c} by employing the so-called {\it Near-Identity Transformation} (NIT) of the dependent variable $\eta(x,t).$ 
The transformation generates a HKdV with coefficients, different from the coefficients of the original HKdV, however, the transformed equation can be matched to (one of) the three known integrable HKdV equations, whose coefficients have particular values.

Let us suppose that $\eta(x,t)$ satisfies the equation 
\begin{align}
    \eta_t+&c\eta_x+\varepsilon A \eta \eta_x + \varepsilon B \eta_{xxx} \nonumber \\
    &+\varepsilon^2 M \eta^2 \eta_x 
    +\varepsilon^2 Q \eta_{5x}+\varepsilon^2(N_1 \eta_x \eta_{xx}+N_2 \eta \eta_{xxx})=0
    \label{GKdV5}
\end{align}
for some constants $c,A,B,M,Q,N_1,N_2,$ which is a general form of equation \eqref{KdV_c}. Let us consider the following NIT relating $\eta$ of  \eqref{GKdV5} to another function $E(x,t):$ 
\begin{equation} \label{NIT}
    \eta(x,t)=E+\varepsilon( a_1E^2 + a_2 E_{xx}+a_3 E_x \partial_x^{-1}E),
\end{equation}
where $a_i$ are 3 constant parameters and the inverse differentiation means integration. This transformation is also known as the Kodama transform \cite{Kod1,Kod}, and appears in previous studies like \cite{DGH,ZhiSig}.
From \eqref{NIT} we obtain by differentiation
\begin{equation}
    \eta_t + c \eta_x  = E_t+ c E_x  + \mathcal{O}(\varepsilon)
\end{equation}
and as far as obviously from  \eqref{GKdV5} $ \eta_t + c \eta_x  =  \mathcal{O}(\varepsilon),$ then
\begin{equation} \label{E0}
     E_t+ c E_x =  \mathcal{O}(\varepsilon).
\end{equation}
Again, from  \eqref{NIT} and using in addition \eqref{E0} we have 
\begin{equation}
    \eta_t + c \eta_x + \varepsilon A \eta \eta_x + \varepsilon B \eta_{xxx}  = E_t+ c E_x +  \varepsilon A E E_x  + \varepsilon B E_{xxx}+ \mathcal{O}(\varepsilon^2)
\end{equation}
and since from  \eqref{GKdV5} $ \eta_t + c \eta_x  + \varepsilon A \eta \eta_x + \varepsilon B \eta_{xxx} =  \mathcal{O}(\varepsilon^2),$ then
\begin{equation}\label{xxx}
     E_t+ c E_x + \varepsilon A E E_x  + \varepsilon B E_{xxx}=  \mathcal{O}(\varepsilon^2).
\end{equation}

In other words, the NIT does not change the original equation up to the terms of order $\varepsilon.$
 
If \eqref{NIT} is applied to the full equation \eqref{GKdV5}, then after some similar straightforward calculations taking into account \eqref{xxx}, one can verify that up to terms of order $\varepsilon^2$ the associated evolution equation for $E$ is 
\begin{align}
    E_t+&c E_x+\varepsilon A E E_x + \varepsilon B E_{xxx} \notag \\
    &+\varepsilon^2 M' E^2 E_x +\varepsilon^2 Q' E_{5x}+\varepsilon^2(N'_1 E_x E_{xx}+N'_2 E E_{xxx})= \mathcal{O}(\varepsilon^3),
    \label{GKdV5E}
\end{align} 
where the $\mathcal{O}(\varepsilon^2) $ terms have the following coefficients (all terms involving $\partial_x ^{-1}E$  miraculously cancelling out):
\begin{align}
    M'&=M+A\left(a_1+\frac{1}{2}a_3\right), \label{m}\\
    Q'&=Q, \\
    N_1'&=N_1+6Ba_1-2Aa_2+3Ba_3, \label{n1}\\
    N_2'&=N_2+3Ba_3. \label{n2}
\end{align}

The transformation \eqref{NIT} could be used for example to relate the solutions of the non-integrable HKdV equation \eqref{GKdV5} for the physical variable $\eta$ to the solution $E(x,t)$ of some integrable equation. Integrable equations have the advantage of possessing so-called soliton solutions, which are usually stable (in time) solitary waves, they interact elastically and recover their initial shape after interaction.  The soliton solutions can be obtained explicitly by various methods, such as the inverse scattering method, \cite{ZMNP}.

The relation of \eqref{KdV_c} to the three known integrable equations of HKdV type can be shown as follows. The known integrable HKdV-type equations are 
\begin{equation}\label{64}
    E'_{\tau}+E'_{5x'}+2(6b+1) E'_{x'}E'_{x'x'}+4(b+1) E' E'_{3x'}+20b (E')^2 E'_{x'}=0,
\end{equation}
where $b=3/2$ corresponds to the second equation from the KdV integrable hierarchy, $b=4$ and $b=1/4$  are the other two integrable cases, known as the Kaup-Kuperschmidt (KK) equation \cite{KK} ($b=4$) and 
Sawada-Kotera equation \cite{SK} ($b=1/4$), which appears also in \cite{CDG,KK}. The soliton solutions 
of the KK equation are obtained in \cite{VSG}. The classification appears in \cite{Mikh} on p. 170, where the two equations $b=4$ and $b=1/4$ are given, the $b=3/2$ one is from the KdV hierarchy and is also a symmetry of the KdV equation - it is mentioned on p. 117.

Equation \eqref{64} can also be rewritten in several equivalent forms. With a Galilean transformation a linear $C_1 E'_{x'}$ term could be generated. The shift $E'\to E'+C_2$ (where $C_1, C_2$ are arbitrary constants), leads to another form of this integrable family of equations, with a new time-like variable $t',$ see the details in \cite{I07}:
\begin{align}\label{abo}
     E'_{t'}&+(C_1 +20bC_2^2) E'_{x'}+4(b+1)C_2 E'_{3x'}+40bC_2 E' E'_{x'} \nonumber \\
    &+E'_{5x'}+2(6b+1) E'_{x'}E'_{x'x'}+4(b+1) E' E'_{3x'}+20b (E')^2 E'_{x'}=0.
\end{align}
Following the re-scaling $E'=\varepsilon \kappa E,$ and $$x'=\frac{x}{\sqrt{\varepsilon}\vartheta},\quad t'=\frac{t}{\sqrt{\varepsilon}\vartheta},$$ and introducing a new constant $c'=C_1 +20bC_2^2$ the above equation \eqref{abo} becomes (see\cite{I07} for details)
\begin{align}
    E_{t}&+c' E_{x}+\varepsilon 4(b+1) C_2 \vartheta^2 E_{3x}+\varepsilon  40bC_2 \kappa E E_{x} \nonumber \\
    &+\varepsilon ^2 (\vartheta ^4 E_{5x}+2(6b+1) \vartheta^2 \kappa E_{x}E_{xx}+4(b+1)\vartheta^2 \kappa E E_{3x}+20b \kappa^2 E^2 E_{x})=0. \label{kdv5scale}
\end{align}
In order for the coefficients of \eqref{GKdV5E} to match the coefficients of the integrable equation \eqref{kdv5scale} we require
\begin{align}
        c'&=c, \\
   4(b+1) C_2 \vartheta^2  &= B, \\
     40bC_2 \kappa &=A , \\
        \vartheta^4&=Q'=Q,  
        \end{align}
In so doing, we obtain 
        \begin{align}
        \vartheta=Q^{1/4}, \quad    C_2=\frac{B}{4(b+1)\sqrt{Q}}, \quad \kappa=\frac{A}{40 b C_2}=\frac{(b+1)A\sqrt{Q}}{10b B},
        \end{align}
  with the remaining matching conditions \eqref{m} -- \eqref{n2} determining the constants $a_i:$ 
    \begin{align}
     4(b+1)\vartheta^2 \kappa&=N_2'=N_2+3Ba_3 ,\quad     a_3=\frac{2(b+1)^2AQ}{15bB^2}-\frac{N_2}{3B},
     \end{align}
  Consequently, we obtain  
     \begin{align}
        N_2'=\frac{2(b+1)^2AQ}{5bB}. \label{N2p}
      \end{align}
      For the remaining coefficients we obtain
      \begin{align}
          20b\kappa^2 &=M'=M+Aa_1+\frac{1}{2} Aa_3, \\
           a_1&=\frac{4(b + 1)^2 A^2Q + 5bA B N_2 - 30b M B^2}{30bAB^2},\\
           M'&=\frac{(b + 1)^2 A^2 Q}{5bB^2},\\
    2(6b+1)\vartheta^2 \kappa&=N_1'=N_1+ 6Ba_1-2Aa_2+3Ba_3 , \\
    \quad a_2&=\frac{(b + 1)QA^2 + b N_1 B A - 6 b M B^2}{2 B b A^2} ,\\
    N_1'&=\frac{(6 b + 1)(b + 1)QA}{5 b B}. \label{endeq}
    \end{align}

Thus we have achieved the following. The non-integrable physical model is given by equation \eqref{KdV_c} with coefficients 
\begin{align}
    A &= \frac{3 c}{2 h}, \,\, B=\frac{c h^2}{6}, \,\, Q = \frac{19 c h^4}{360}, \,\,
M =-\frac{3 c}{8 h^2}, \,\, N_1 = \frac{23 c h}{24}, \,\, N_2 =\frac{5 c h}{12}.
\end{align}

By employing a NIT \eqref{NIT} this equation is transformed to an integrable one \eqref{GKdV5E}
with coefficients given by \eqref{N2p} -- \eqref{endeq}: 
\begin{align}
     Q'&=Q=\frac{19 c h^4}{360}, \,\, M'=\frac{171 (b + 1)^2 c}{200 b h^2},\\
   N_1'&=\frac{19(6b + 1)(b + 1)ch}{200  b} ,\,\, N_2'=\frac{19(b + 1)^2 ch}{100 b}.
\end{align}
Moreover, the parameters of the NIT \eqref{NIT} are
\begin{align}
    a_1&= \frac{57 b^2 + 214 b + 57}{150 b h}, \,\, a_2=\frac{(202b + 57)h^2}{360 b }, \,\, 
    a_3=\frac{57 b^2 - 11b + 57}{150  b h }.
\end{align}
We note that both the coefficients and the NIT parameters depend on the parameter $b,$  and the three known integrable equations of this type correspond to the three possible values: $b=3/2,$ $b=4$ and $b=1/4.$ 

\section{Discussion}

In this chapter we have presented in detail the procedure of the derivation of the higher order KdV model (HKdV) for surface waves and its relation to three integrable equations. These integrable models are solvable by the inverse scattering method and possess soliton solutions. Another interesting problem concerning the HKdV equation is its connection to the Camassa-Holm-type equations, however this has not been explored in the present contribution. The methods illustrated here can be extended in different directions, for example, for derivation of long-wave models for internal waves.
The zero vorticity assumption is essential since in this case the surface dynamics allows an analytic continuation to the fluid domain. The results could possibly be extended in a similar way to the case of constant vorticity, and this is work in progress. Arbitrary nonzero vorticity however leads to complicated interactions between the physical quantities at surface and in the fluid volume, see for example \cite{IKI23}. 

\subsection*{Acknowledgements} This publication has emanated from research conducted with the financial support of Science Foundation Ireland under Grant number 21/FFP-A/9150, and from discussions, undertaken at the workshop {\it Nonlinear Dispersive Waves} held at University College Cork in April 2023.  


\end{document}